\documentclass[%
reprint,
showpacs,preprintnumbers,
 amsmath,amssymb,
 aip,
 graphicx,
 jap,
]{revtex4-1}

\usepackage{graphicx}
\usepackage{dcolumn}
\usepackage{bm}
\usepackage[mathlines]{lineno}

\begin{document}
\preprint{}

\title{Control of carrier transport in GaAs by longitudinal-optical phonon-carrier scattering using a pair of laser pump pulses}
\author{Muneaki Hase}
\email{mhase@bk.tsukuba.ac.jp}
\affiliation{Institute of Applied Physics, University of Tsukuba, 1-1-1 Tennodai, Tsukuba 305-8573, Japan}
\author{Daisuke Hayashi}
\affiliation{Institute of Applied Physics, University of Tsukuba, 1-1-1 Tennodai, Tsukuba 305-8573, Japan}
\author{J. D. Lee}
\affiliation{School of Materials Science, Japan Advanced Institute of Science and Technology, Ishikawa 923-1292, Japan}

\date{\today}

\begin{abstract}
We demonstrate optical control of the LO phonon-plasmon coupled (LOPC) modes in GaAs by using a femtosecond 
pump-pulse pair. 
The relaxation time of the plasmon-like LOPC mode significantly depends on the separation time ($\Delta t$) of the 
pump-pulse pair.  Especially it is maximized when $\Delta t$ becomes simultaneously comparable to the half period of 
the longitudinal optical (LO) phonon oscillation and resonant to the 3/4 period of the plasmon-like LOPC oscillation. We attribute these 
observations to the modification of carrier-LO phonon scattering and ballistic motion of the plasmon-like LOPC mode. 
\end{abstract}

\pacs{78.47.J-, 72.30.+q, 63.20.kd, 72.80.Ey}

\keywords{coherent phonon, phonon-plasmon interactions, optical pumping}

\maketitle

\section{INTRODUCTION}

Ballistic transport of carriers in semiconductor devices, such as high electron mobility transistor (HEMT) and 
field-effect transistor (FET), possesses a potential issue to realize terahertz (10$^{12}$ Hz) ultra-high-speed 
operation in telecommunication and optical memory. To date intriguing efforts have been made to realize the 
ballistic regime in semiconductors,\cite{Wang03,Glinka08} carbon nanotubes,\cite{Chen05} and 
graphene.\cite{Du08}  These studies in semiconductor nanostructures basically rely on the device structures to 
have the ballistic condition $\ell > L$, where $\ell$ is the carrier mean free path length and $L$ is the length of the 
nanostructure,\cite{Wang03} while ballistic transport of carriers in graphene would be governed by ambipolar 
electric field effect.\cite{Geim07}

Since mobility ($\mu$) of carrier transport in semiconductors is governed by inelastic scattering processes, such as electron-hole 
scattering, electron-phonon scattering, and scattering by defects and disorders,\cite{Yu99} another direction to 
realize the ballistic regime in semiconductor is to manipulate those scattering processes in semiconductors without 
making nanostructure. 
In fact, the ballistic regime for plasma wave propagation has been examined in a bulk GaAs,\cite{Cummings01,Glinka08}  
and in metal surfaces \cite{Lisowski04} after the excitation of electron-hole plasma by femtosecond laser pulses, 
in which reduction of the electron-phonon scattering was significantly suggested. In a polar semiconductor, such as 
GaAs and InP, Fr\"{o}hlich-type carrier-LO phonon scattering plays a major role at room temperature.\cite{Yu99,Kersting98} 

In doped semiconductors, such as $n$-type GaAs, it is well known that the 
plasmon and the longitudinal optical (LO) phonon form a new quasiparticle, the LO phonon-plasmon coupled (LOPC) mode, 
through Coulomb interactions.\cite{Mooradian66} 
At the high doping levels ($n_{dop}$ $\geq$ 1.0$\times$10$^{18}$cm$^{-3}$) the upper branch of the LOPC mode becomes plasmon-like, whose relaxation 
is dominated by the decay of carrier plasma.\cite{Vallee97,Hase09}  

One of the potential applications of femtosecond laser pulse sequences is controlling the amplitudes of coherent 
collective motions of atoms excited in condensed matters.\cite{Weiner90,Dekorsy93,Hase96} One can enhance 
the amplitude of a phonon mode by applying $in-phase$ pulses, or suppress it by applying $out-of-phase$ pulses, 
both of which are observed in real time-domain. 
Recently, it has been theoretically pointed out that coherent control of relaxation of carrier plasma is possible by the 
irradiation of THz-rate trains of laser pulses with {\it below-gap} excitation condition,\cite{Lee08,Lee09} while most 
of the pump-probe experiments and time-domain simulations have been made with {\it above-gap} excitation in 
GaAs,\cite{Saeta91,Cho96,Hase98L,Cummings01,Chang02a,Dumitrica02} 
where the pump-pulse generates substantial nonequilibrium carriers in the conduction band with excess energy. 
In this condition, one expects that the firstly generated coherent LOPC modes may be always suppressed by the 
secondly pump-pulse if the separation time between the pump-pulses is several hundred femtoseconds \cite{Bakunov03}.  

In this paper, we report experimentally on the coherent control of the relaxation time of the plasma-relevant coherent 
LO phonon-plasmon coupled mode using a femtosecond pump-pulse pair in $n$-GaAs under nearly {\it below-gap} excitation 
condition. 
The relaxation time of the plasma-relevant LOPC mode is directly linked to the carrier transport,\cite{Lee08,Lee09,Hase09} and 
therefore, our results are suggestive to a possibility of realization of near-ballistic regime as a consequence of the 
reduction of the electron-phonon scattering. 

\begin{figure}
\includegraphics[width=8.0cm]{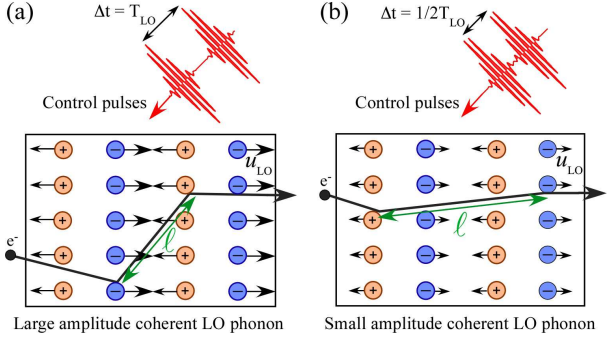}
\caption{(Color online). (a) Schematic of the coherent control of the amplitude of the LO phonon 
and subsequent manipulation of the carrier transport in the case of the large amplitude LO 
phonon, in which the carrier mean free path length $\ell$ satisfies  $\ell \approx \ell_{th}$, with $\ell_{th}$ the thermal 
mean free path length. (b) The case of the small amplitude LO phonon, in which $\ell$ satisfies $\ell \geq \ell_{th}$. 
$T_{LO}$ is the time period of the LO phonon.  
}
\label{Fig1}
\end{figure}

\section{EXPERIMENTAL TECHNIQUE}

To detect the plasmon-like LOPC modes in real time-domain, we used an electro-optic (EO) 
detection \cite{Cho90} combined 
with a fast scan pump-probe method utilizing a mode-locked Ti:sapphire laser with a pulse 
duration of 20 fs, and a center wavelength of 855 nm (= 1.45 eV). 
The femtosecond pump-pulse pair was generated through a Michelson-type interferometer, in which a 
closed-loop piezo stage was installed to adjust the time separation ($\Delta t$) of the two pump-pulses to 
a precision of $<$ 0.1 fs. 
The average power of each pump-beam was fixed at 30 mW, while that 
of the probe was kept at 3 mW. The maximum photo-excited carrier density by the each pump-
pulse was estimated to be $n_{exc}$=4$\times$10$^{17}$ cm$^{-3}$ from the pump power 
density (60 $\mu$J/cm$^{2}$) and the absorption coefficient. 
The sample used was Si-doped $n$-type GaAs with a doped carrier density of 
$n_{dop}$= 1.0 $\pm$ 0.3 $\times$10$^{18}$cm$^{-3}$, which is larger than $n_{exc}$.
Upon the doped carrier concentration of $n_{dop}$= 1.0$\times$10$^{18}$cm$^{-3}$, 
the Burstein-Moss effect would shift the band-gap energy in $n$-GaAs to higher energy of 
$\approx$ 1.45-1.48 eV from the original value (1.43 eV at 295 K).\cite{Borghs89} 
By this effect, our laser can stand at nearly {\it below-gap} excitation condition, although 
the phase-breaking scattering
of virtual excitations can create the Urbach tail in the absorption line of an order of several 
meV,\cite{Leitenstorfer94} generating small free-streaming current in the depletion region. 
The transient EO reflectivity change 
($\Delta R_{eo}/R$) was recorded as a function of the time delay ($\tau$) at room temperature (295 K). 

\section{EXPERIMENTAL RESULTS AND ANALYSIS}

\begin{figure}
\includegraphics[width=8.0cm]{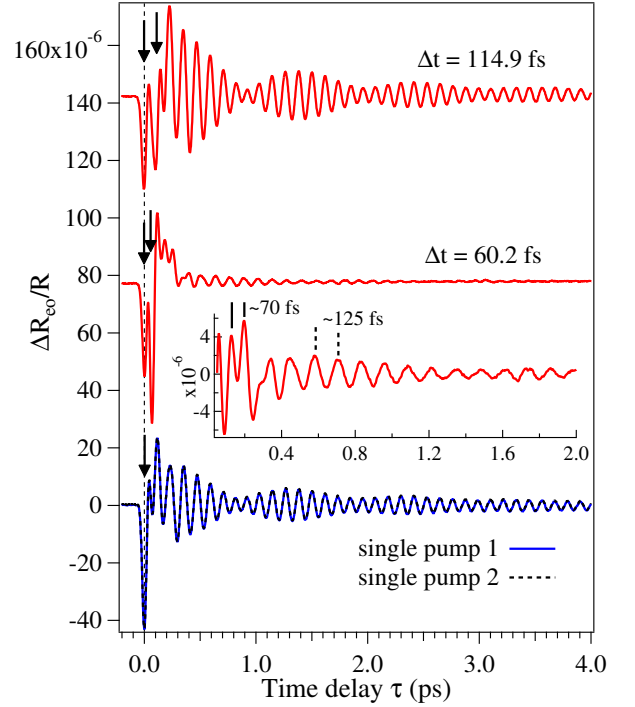}
\caption{(Color online). The $\Delta R_{eo}/R$ signals for $n$-GaAs obtained by each single pump 
(bottom) and pump-pulse pairs with two different $\Delta t$s. The inset shows 
the enlargement of the oscillation with $\Delta t$ = 60.2 fs after subtraction of the exponentially decaying electronic background. 
The vertical arrows represent the temporal position of the pump-pulses. 
}
\label{Fig2}
\end{figure}

As shown in Fig. 2, the signal observed in $n$-GaAs with the single pump-pulse exhibits 
a beating oscillation, which has been commonly obtained by the similar pump-probe 
experiments \cite{Cho96,Hase98L,Chang02a} and assigned to the coexistence of the LO phonon 
and the lower branch (LOPC-) of the LOPC modes. The coherent LO phonon is excited in the depletion region 
($d$ = 0 $-$ 100 nm), while the LOPC modes are excited in both depletion and bulk region.\cite{Cho96,Hase98L,Chang02a} 
When the sample is excited with pump-pulse pair, the constructive and destructive generation of 
coherent phonons is observed. Especially, a high frequency coherent oscillation (see the inset of Fig. 2) 
becomes evident up to $\tau$ $\sim$ 300 fs when $\Delta t$ is 60.2 fs, whose oscillation period is 
$\approx$ 70 fs (= 14.3 THz). This mode appears to be the higher branch of the LOPC 
mode (LOPC+) as seen in the corresponding Fourier transformed (FT) spectra, Fig. 3, which is 
converted from the time-domain data just after the second pump-pulse excitation to exclude any 
coherent artifact originating from the period of the pump-pulse pair. The almost complete 
destructive interference of the coherent LO phonon observed in Fig. 3 ($\Delta t$ = 60.2 fs) may imply generating 
a similar condition to phonon cooling near the surface of the sample.\cite{Bahae04} 
As shown in the inset of Fig. 3, the frequencies of the LOPC modes are in good agreement with the 
calculation within the experimental errors 
under the condition that the total carrier density $n_{t}$ is the sum of the doped- and photoexcited-carriers 
($n_{t} = n_{dop} + n_{exc}$).\cite{Cho96,Hase98L} 

\begin{figure}
\includegraphics[width=8.0cm]{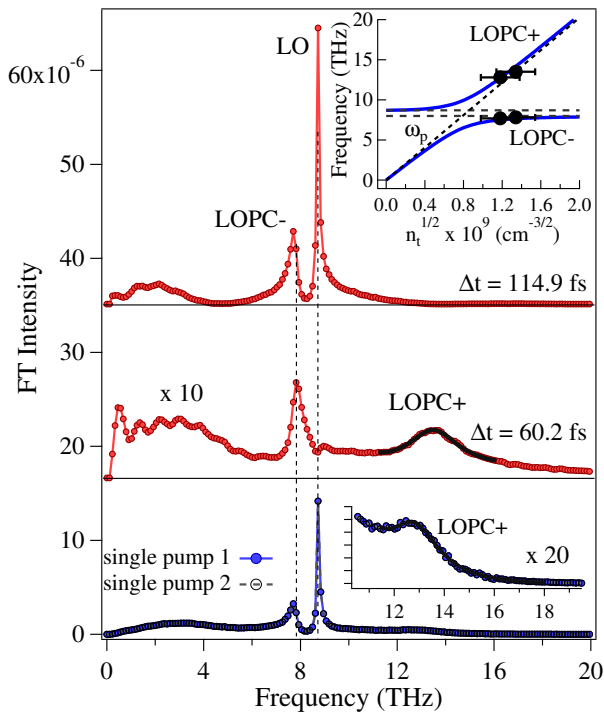} 
\caption{(Color online). The FT spectra obtained from the time-domain data in Fig. 2. 
With single pump excitation, the LO phonon (8.8 THz), the LOPC- mode (7.8 THz), and the very weak LOPC+ mode 
(13 THz) are observed. 
The upper inset describes LOPC frequencies as a function of the total carrier density $n_{t}$. The solid curves 
are the frequencies of the LOPC+ and LOPC- modes calculated by the equilibrium model of Ref. [10]. 
The horizontal dashed lines are the frequencies of the bare LO (8.8 THz) and TO (8.0 THz) phonons, while the 
plasma frequency ($\omega_{p}$) depends linearly on $n_{t}^{1/2}$. The black thick lines around the 
LOPC+ mode represent the fit of the data with the Lorenz function. 
}
\label{Fig3}
\end{figure}

Careful curve fitting with a Lorenz function to the FT data at various $\Delta t$s enables us to estimate the 
relaxation time of the LOPC+ mode by using the relation between the linewidth and the relaxation time,\cite{Laubereau78,Hase98} that changes from 180 $\pm$ 20 to 235 $\pm$ 20 fs, depending 
on the value of $\Delta t$. 
The relaxation time of the coherent LOPC+ mode in highly doped $n$-GaAs obtained under the {\it above-gap} excitation 
was 135 $\pm$ 15 fs at $n_{t}$ $\approx$ 1.1$\times$10$^{18}$cm$^{-3}$, \cite{Hase09} and 130 $\pm$ 40 fs at 
$n_{t}$ $\approx$ 1.2$\times$10$^{18}$cm$^{-3}$. \cite{Hase99}  Comparing these values with the relaxation time 
obtained in the present study indicates that the nearly {\it below-gap} excitation makes the relaxation slightly longer. 

\begin{figure}
\includegraphics[width=8.8cm]{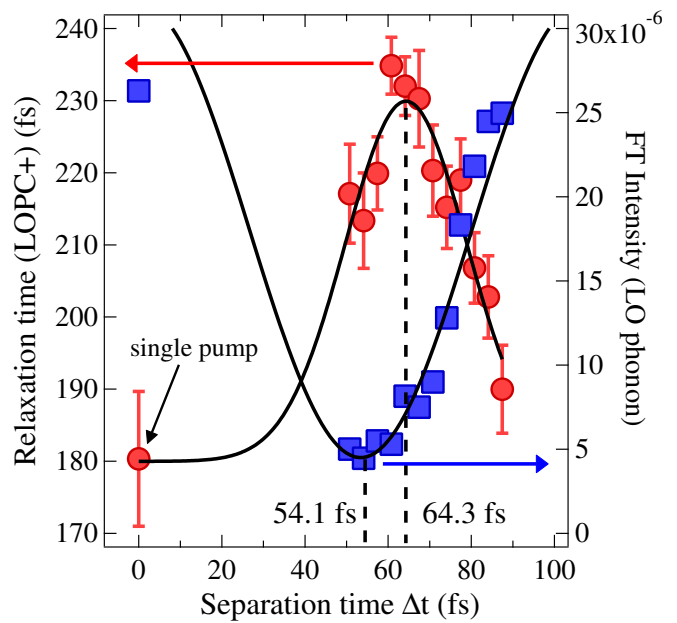} 
\caption{(Color online). Relaxation time of the plasmon-like LOPC+ mode together with the peak intensity of the LO 
phonon as a function of the separation time $\Delta t$. The LOPC+ mode cannot be separately observed when 
$\Delta t$ $\leq$ 40 fs, and $\Delta t$ $\geq$ 90 fs due to the stronger spectral tail by the LO mode. The thick 
curve corresponds to the fit of the data with a Gaussian function to estimate the maximum position of 
the relaxation time, while that for the LO intensity is a guide for the eyes with a cosine function. 
}
\label{Fig4}
\end{figure}

Figure 4 plots the relaxation time of the plasmon-like LOPC+ mode together with the FT intensity of 
the coherent LO phonon as the function of the separation time $\Delta t$. The relaxation time exhibits 
a maximum at $\Delta t$ $\approx$ 64 fs, while the amplitude of the LO phonon shows minimum at 
$\Delta t$ $\approx$ 54 fs. The latter effect can be understood in terms of the destructive interference 
of the coherent LO phonon as observed in GaAs \cite{Dekorsy93} and other materials.\cite{Hase96} 
The value of  $\Delta t$ $\approx$ 64 fs is significantly close to the characteristic time of (3/4)$T_{LOPC+}$ 
(= 57.8 fs), at which the LOPC+ oscillation reaches its maximum (when assuming the LOPC+ mode is a sine-like oscillation). 
To investigate the oscillatory phase of the LOPC+ mode, 
we fit the time-domain data of the single pump [labeled 1; bottom in Fig. 2] with damped harmonic oscillations (not shown), 
and as the results, the sine-like phase was confirmed for the LOPC+ and LOPC- modes. On the contrary, the 
cosine-like phase was observed for the LO mode, being consistent with the previous result.\cite{Pfeifer92} 
According to Ref. [23], if $\Delta t$ does not match a maximum of the oscillation of the LOPC modes, 
the oscillation is suppressed and the ballistic motion cannot be obtained. 
On the contrary, $\Delta t$ matching a maximum of the oscillation would give long-surviving motion, suggesting 
$\Delta t$ = (3/4)$T_{LOPC+}$ should be a necessary condition for ballistic motion of the LOPC+ mode. 
Thus, the enhancement of the relaxation time at $\Delta t$ $\approx$ 64 fs satisfies approximately both 
conditions; (i) minimum  LO phonon amplitude 
($\Delta t$ $\approx$ $T_{LO}$/2), and (ii) ballistic motion of the LOPC+ mode by coherent excitation with 
$\Delta t$ $\approx$ (3/4)$T_{LOPC+}$. 

\section{DISCUSSION}

A plausible reason for the enhancement of the relaxation time is reduction of the Fr\"{o}hlich-type carrier-LO phonon 
scattering due to minimizing the LO phonon amplitude, which would occur at the value of $\Delta t$ $\approx$ 57.5 fs (=$T_{LO}$/2) 
(see Fig. 1). At room temperature (295 K), however, there is a substantial thermal population of LO phonons, which 
interact with the plasmon-like LOPC mode and would contribute to its relaxation. This contribution might be much 
stronger than the contribution of the coherent LO phonon generated and manipulated optically. In order to check these 
possibilities, we consider the ratio of the total number of the coherent LO phonons $n_{coh}$ to the total number of the 
LO phonons as given by,\cite{Nakayama07} 
\begin{equation}
A(T) = \frac{n_{coh}}{n_{coh} + n_{incoh}(T)},
\end{equation}
where $n_{incoh}(T)$ is the number of thermal (incoherent) LO phonons expressed by 
$n_{incoh}(T) = n_{0}/[exp(\hbar \omega_{LO}/k_{B}T) - 1]^{-1}$. In the case of the GaAs-like LO phonon the ratio $n_{coh}/n_{0}$ 
was experimentally obtained to be 0.091 at 1 $\mu$J/cm$^{2}$.\cite{Nakayama07} Consequently we would expect  $n_{coh}/n_{0}$  
$\approx$ 1 (maximum coherence) in our case (60 $\mu$J/cm$^{2}$). Then we found from Eq.  (1) that the ratio of the coherent LO 
phonon is  $A$(T = 295 K) $\approx$ 0.78. It follows that the ratio of the thermal LO phonon is 1- $A$(T = 295 K) $\approx$ 0.22. 
Although there is the constant thermal population of LO phonons as estimated above, the contribution of the coherent LO phonon 
is enough to affect the relaxation of the plasmon-like LOPC mode under our experimental condition. 

Based on the above investigations, we argue overall understanding of the enhancement of the relaxation time of the 
LOPC+ mode in more details. In the case of the thermal LO phonon, high enough temperatures enable all LO phonons to participate 
in the Fr\"{o}hlich-type carrier-LO phonon scattering process, which involves the absorption and emission of LO phonons; the interaction 
occurs very often at room temperature, and then reduces the mobility.\cite{Yu99,Kersting98} To link this scenario with our case, we are controlling 
the amplitude of the coherent LO phonon [ $A$(T = 295 K) $\approx$ 78 \% in the maximum] by the optical means, instead of changing the 
lattice temperature, leading to the reduction of the probability of scattering events if the amplitude of the coherent LO phonon becomes 
minimum, where  $A$(T = 295 K) $\approx$ 0. Additionally, the synchronization of free-streaming current by the second pump-pulse 
would lead to the condition of another ballistic motion of plasmons, leading to the subsequent formation of the ballistic motion of the 
plasmon-like LOPC+ mode, whose oscillation persists up to $\tau$ $\sim$ 300 fs, as uncovered in the inset of Fig. 2. 

Note that other scattering processes than the Fr\"{o}hlich-type carrier-phonon 
scattering cannot contribute to the "$\Delta t$ dependence" at room temperature because 
the arrival time of the second pump-pulse changes only the amplitude of the LO phonon, and 
not change carrier density, defect density, and amplitude of acoustic phonons within the 
time scale of a few hundred femtoseconds. 

\section{CONCLUSIONS}

To summarize, we have obtained the enhancement of the relaxation time of the plasmon-like LOPC+ mode
through the coherent control of the amplitude of both the LO phonon and plasmon-like LOPC 
modes in n-GaAs by use of the femtosecond pump-pulse pair. 
This effect would be attributed to the two possible origins, that is the minimum Fr\"{o}hlich-type carrier-LO phonon 
scattering and ballistic motion of the plasmon-like LOPC+ mode, induced by coherent excitation with the ultimate condition of 
$\Delta t$ $\approx$ $T_{LO}$/2 $\approx$ (3/4)$T_{LOPC+}$. 

\begin{acknowledgments}
This work was supported in part by PRESTO-JST, KAKENHI-22340076 and 
Special Coordination Funds for Promoting Science and Technology from MEXT, Japan. 
\end{acknowledgments}


\begin{thebibliography}{99}
 
\bibitem{Wang03}
J. Wang and M. Lundstrom, IEEE Trans. Electron Devices {\bf 50}, 1604 (2003).
  
 \bibitem{Glinka08}
Y. D. Glinka, D. Maryenko, and J. H. Smet, Phys. Rev. B {\bf 78}, 035328 (2008). 
  
 \bibitem{Chen05}
 Y. -F. Chen and M. S. Fuhrer, Phys. Rev. Lett. {\bf 95}, 236803 (2005).

\bibitem{Du08}
X. Du, I. Skachko, A. Barker and E. Y. Andrei, Nat. Nanotech. {\bf 3}, 491 (2008).

\bibitem{Geim07}
A. K. Geim and K. S. Novoselov, Nat. Mat. {\bf 6}, 183 (2007).

\bibitem{Yu99}
P. Y. Yu and M. Cardona, Chap. 3 in {\it Fundamentals of Semiconductors},
 (Springer-Verlag, Berlin Heidelberg 1999).
 
\bibitem{Cummings01}
M. D. Cummings, J. F. Holzman, and A. Y. Elezzabi, Appl. Phys. Lett. {\bf 78}, 3535 (2001). 

\bibitem{Lisowski04}
M. Lisowski, P. A. Loukakos, U. Bovensiepen, J. St\"{a}hler, C. Gahl, and M. Wolf, 
Appl. Phys. A, {\bf 78}, 165 (2004). 

\bibitem{Kersting98}
R. Kersting, J. N. Heyman, G. Strasser, and K. Unterrainer, Phys. Rev. B {\bf 58}, 4553 (1998). 

\bibitem{Mooradian66}
A. Mooradian and G. B. Wright, Phys. Rev. Lett. {\bf 16}, 999 (1966).

\bibitem{Vallee97}
F. Vall\'{e}e, F. Ganikhanov, and F. Bogani, Phys. Rev. B {\bf 56}, 13141 (1997). 

\bibitem {Hase09}
M. Hase, Appl. Phys. Lett. \textbf{94}, 112111 (2009). 

\bibitem{Weiner90}
A. M. Weiner, D. E. Leaird, G. P. Wiederrecht, and K. A. Nelson, Science {\bf 247}, 1317 (1990). 

\bibitem{Dekorsy93}
T. Dekorsy, W. A. K\"{u}tt, T. Pfeifer, and H. Kurz, Europhys. Lett. {\bf 23}, 223 (1993). 

\bibitem{Hase96}
M. Hase, K. Mizoguchi, H. Harima, S. Nakashima, M. Tani, K. Sakai, and M. Hangyo, 
Appl. Phys. Lett. {\bf 69}, 2474 (1996).

\bibitem{Lee08}
J. D. Lee and M. Hase, Phys. Rev. Lett. {\bf 101}, 235501 (2008). 

\bibitem{Lee09}
J. D. Lee, H. Gomi, and M. Hase, J. Appl. Phys. {\bf 106}, 083501 (2009).

\bibitem{Saeta91}
P. Saeta, J.-K. Wang, Y. Siegal, N. Bloembergen, and E. Mazur, Phys. Rev. Lett. {\bf 67}, 1023 (1991). 

\bibitem{Cho96}
G. C. Cho, T. Dekorsy, H. J. Bakker, R. H\"{o}vel, and H. Kurz, Phys. Rev. Lett. {\bf 77}, 4062 (1996). 

\bibitem{Hase98L}
M. Hase, K. Mizoguchi, H. Harima, F. Miyamaru, S. Nakashima, R. Fukasawa, M. Tani, and K. Sakai, 
J. Luminescence {\bf 76-77}, 68 (1998).

\bibitem{Chang02a}
Y. -M. Chang, Appl. Phys. Lett. {\bf 80}, 2487 (2002). 

\bibitem{Dumitrica02}
T. Dumitric\'{a} and R. E. Allen, Phys. Rev. B {\bf 66}, 081202(R) (2002).

\bibitem{Bakunov03} 
M. I. Bakunov, A. V. Maslov, and S. N. Zhukov, Phys. Rev. B {\bf 67}, 153201 (2003).

\bibitem{Cho90}
G. C. Cho, W. K\"{u}tt, and H. Kurz, Phys. Rev. Lett. {\bf65}, 764 (1990).
                   
\bibitem{Borghs89}
G. Borghs, K. Bhattacharyya, K. Deneffe, P. Van Mieghem, and R. Mertens, 
J. Appl. Phys. {\bf 66}, 4381 (1989). 

\bibitem{Leitenstorfer94} 
A. Leitenstorfer, A. Lohner, T. Elsaesser, S. Haas, F. Rossi, T. Kuhn, W. Klein, G. Boehm, 
G. Traenkle, and G. Weimann, Phys. Rev. Lett. {\bf 73}, 1687 (1994).

\bibitem{Bahae04} 
M. Sheik-Bahae and R. I. Epstein, Phys. Rev. Lett. {\bf 92}, 247403 (2004). 

\bibitem{Laubereau78} 
A. Laubereau and W. Kaiser, Rev. Mod. Phys. {\bf 50}, 607 (1978).

\bibitem{Hase98}
M. Hase, K. Mizoguchi, H. Harima, S. Nakashima, and K. Sakai, Phys. Rev. B \textbf{58}, 5448 (1998). 

\bibitem{Hase99}
M. Hase, S. Nakashima, K. Mizoguchi, H. Harima, and K. Sakai, Phys. Rev. B {\bf 60}, 16526 (1999).

\bibitem{Pfeifer92} 
T. Pfeifer, T. Dekorsy, W. A. K\"{u}tt, and H. Kurz, Appl. Phys. A  {\bf 55}, 482 (1992). 

\bibitem{Nakayama07}
M. Nakayama, K. Mizoguchi, O. Kojima, T. Furuichi, A. Mizumoto, S. Saito, A. Syouji, and K. Sakai, 
phys. stat. sol. (a) {\bf 204}, 518 (2007). 

\end{thebibliography}
\end{document}